# H$_2$ Temperatures in the Crab Nebula

E. D. Loh,[1,*] J. A. Baldwin,[1,†] G. J. Ferland,[2,‡] Z. K. Curtis,[1,§] , C. T. Richardson,[1,**] A. C. Fabian[3,††] and Philippe Salomé[4,‡‡]

[1] Department of Physics and Astronomy, Michigan State University, East Lansing, MI 48824-2320 USA
[2] Department of Physics, University of Kentucky, Lexington, KY 40506, USA
[3] Institute of Astronomy, University of Cambridge, Madingley Road, Cambridge CB3 0HA
[4] LERMA & UMR8112 du CNRS, Observatoire de Paris, 61 Av. de l'Observatoire, F-75014 Paris, France

## Abstract

We used K-band spectra to measure the H$_2$ excitation temperatures in six molecular knots associated with the filaments in the Crab Nebula. The temperatures are quite high – in the range $T \sim 2000–3000$K, just below the H$_2$ dissociation temperature. This is the temperature range over which the H$_2$ 1-0 S(1) line at $\lambda 2.121 \mu$m has its maximum emissivity per unit mass, so there may be many additional H$_2$ cores with lower temperatures that are too faint to detect. We also measured the electron density in adjacent ionized gas, which on the assumption of gas pressure balance indicates densities in the molecular region $n_{mol} \sim 20,000$ H baryons cm$^{-3}$, although this really is just a lower limit since the H$_2$ gas may be confined by other means. The excited region may be just a thin skin on a much more extensive blob of molecular gas that does not have the correct temperature and density to be as easily detectable. At the opposite extreme, the observed knots could consist of a fine mist of molecular gas in which we are detecting essentially all of the H$_2$. Future CO observations could distinguish between these two cases. The Crab filaments serve as the nearby laboratories for understanding the very much larger filamentary structures that have formed in the intracluster medium of cool-core galaxy clusters.

[*] loh@msu.edu

[†] baldwin@pa.msu.edu

[‡] gary@pa.uky.edu

[§] curtisza@msu.edu

[**] richa684@msu.edu

[††] acf@ast.cam.ac.uk

[‡‡] philippe.salome@obspm.fr

# 1. Introduction

The Crab Nebula presents a unique opportunity to study a very young supernova remnant (SNR) at high spatial resolution and at times well before it strikes the interstellar medium. The Crab's extensive filament system is still dominated by the supernova phenomenon itself, rather than by the later mixing and shocks. Because of the importance of SNe in defining the Interstellar Medium (ISM) at high redshift, the molecular and dust inventory of the Crab's filament system is a question of far-reaching importance.

It is surprising that molecules can even exist in an environment as adverse as a young SNR, but Graham, Wright & Longmore (1990; hereafter G90) discovered bright $H_2$ emission in the near infrared. We (Loh, Baldwin & Ferland 2010; Loh et al. 2011; hereafter Papers I & II) have recently carried out a near-infrared survey at considerably higher spatial resolution than G90. We initially identified an especially bright $H_2$ knot (designated Knot 1) and subsequently found another 54 knots with strong emission in the $H_2$ 2.121 µm 1–0 S(1) line. These knots clearly are associated with the system of filaments seen in optical emission lines from ionized gas, and are high surface-brightness clumps set against a background of fainter $H_2$ emission also associated with the filaments (Paper II).

The next step is to quantify the physical conditions in the $H_2$ knots since those conditions determine the chemistry and emission. The $H_2$ emission must be understood in the context in which it occurs. The molecular cores are surrounded by sheaths ionized by the Crab's synchrotron continuum and which form the ionized filamentary structure seen at visible wavelengths (e.g. Sankrit et al. 1998). The observed properties of this ionized sheath surrounding the $H_2$ blobs sets the boundary conditions which any successful model of the $H_2$ region must satisfy. The simplest expectations are clear. The ionized layer extends over the thickness needed to absorb the far-UV synchrotron continuum. There should then be $H_0$ and $H_2$ layers, each successively cooler, as the penetrating non-thermal continuum is attenuated. Fragile molecules will exist in only the deepest, most shielded, regions. In the simplest case constant gas pressure or hydrostatic equilibrium may apply, and cooler regions will have a correspondingly higher density. $H_2$ will exist in regions that are most shielded from the energizing continuum, and so will be very cold. Typical gas kinetic temperatures in regions where most H is $H_2$ are expected to be roughly 100 K. This is too low to effectively excite the $H_2$ levels producing 2 µm emission, which lie some 7000 K above ground, so only very faint $H_2$ emission produced by continuum fluorescent excitation (Black & Dalgarno 1976; van Dishoeck & Black 1986; Sternberg & Dalgarno 1989) should be observed, and it should come from cold gas.

Here we report measurements of the populations of a sufficient number of $H_2$ levels to derive excitation temperatures for several of these knots. We will show that they are quite warm, close to the dissociation temperature of the molecule, which accounts for the high $H_2$ surface brightness.

Warm $H_2$ emission is commonly found in regions where shocks strike molecular clouds in the ISM. Examples include the Becklin-Neugebauer object in Orion (Gautier et al. 1976), merging galaxies (cf. Lutz et al. 2003) and later stages of SNR evolution where the ejecta plows into the ISM (cf. Shinn et al. 2011). However, it is not at all clear whether or not shocks should play a significant role in exciting the Crab filaments, which are at least approximately co-expanding with the surrounding shell.

In their discovery paper G90 discussed two other mechanisms which may warm the $H_2$ region: penetrating x-rays and ionizing particles. Our present study of the Crab filaments was originally motivated by their morphological and spectroscopic similarities to filaments surrounding brightest cluster galaxies in cool-core clusters. These cluster filaments are thought to play an as yet poorly understood role in the feedback process coupling the central AGN with the surrounding intracluster medium. We have shown that ionizing particles are responsible for exciting and ionizing the cool-core cluster filaments (Ferland et al. 2009) and that the hot intracluster gas is the source of these particles (Fabian et al. 2011). The Crab offers a laboratory to study very similar processes at much higher spatial resolution and signal to noise than the extragalactic case. No $H_2$ spectra of a very young SNR such as the Crab, where collisional interactions are not yet clearly dominant, exist in the literature to our knowledge. The infrared spectroscopy described here is the necessary next step towards a fuller understanding of the processes at work in this type of molecular core, with implications for both the effects of SNe on dust and gas abundances in the ISM, and the nature of distant structures such as the cool-core cluster filaments.

## 2. Observations

We obtained near-infrared spectra of seven knots. We used the OSIRIS spectrometer[§§] on the 4m SOAR Telescope[***] on four nights. The spectra were taken through a 0.42 × 72 arcsec slit, covering the wavelength range 2.02–2.33μm with resolution $\lambda/\Delta\lambda = 3000$ and a scale along the slit of 0.139 arcsec/pixel. On the first night we observed the brightest $H_2$ feature, Knot 1, with the slit also crossing the adjacent bright, ionized filament FK-10 (Fesen & Kirshner 1982) and a star that we used to establish the slit position. On the remaining three nights, we worked our way down through the next few brightest knots from the list given in Paper II, with the spectrograph rotated in each case so that the slit would also cover a second knot of intermediate brightness. The targets were observed in a series of 600s exposures interspersed with 150s exposures on sky positions chosen to lie well off the Crab Nebula. The position of the knot along the slit was changed by several arcsec between each successive exposure, to average out small-scale sensitivity patterns on the detector. Table 1 lists some of the observational details for each exposure. The 'slit centre' given in the table is for a point midway between the pair of knots observed on each night,

| Table 1. Observations | | | | |
|---|---|---|---|---|
| Knot | Date (UT) | Slit Centre (J2000 R.A., Dec) | PA (deg) | Total Exp. Time (min) |
| 1 | 2010 Nov 18 | 05:34:34.32 +21:59:49.1 | 178.0 | 130 |
| 38, 40 | 2010 Dec 14 | 05:34:33.22 +22:02:13.6 | 97.5 | 120 |
| 44, 46 | 2011 Jan 18 | 05:34:31.74 +22:01:52.8 | 135.0 | 120 |
| 50, 51 | 2011 Jan 19 | 05:34:27.68 +22:01:50.3 | 139.0 | 180 |

[§§] OSIRIS is a collaborative project between the Ohio State University and Cerro Tololo Inter-American Observatory (CTIO) and was developed through NSF grants AST 90-16112 and AST 92-18449. CTIO is part of the National Optical Astronomy Observatory (NOAO), based in La Serena, Chile. NOAO is operated by the Association of Universities for Research in Astronomy (AURA), Inc. under cooperative agreement with the National Science Foundation.

[***] The Southern Astrophysical Research Telescope is a joint project of Michigan State University, Ministério da Ciência e Tecnologia-Brazil, the University of North Carolina at Chapel Hill, and the National Optical Astronomy Observatory.

The spectra were reduced following typical procedures for long-slit optical spectra: dark frames were subtracted, pixel-by-pixel variations were corrected using a normalized flat made from a quartz lamp, and then vignetting effects were corrected using an illumination frame made from the average of the sky frames. The sky spectra were smoothed with a 5-pixel median filter in the direction along the slit to make up for their shorter exposure times, and then a scaled average of the sky spectra taken just before and after each exposure on the Crab was subtracted from that exposure. The wavelength scale was calibrated using the night sky OH emission lines, and the fluxes were calibrated using three stars each night with accurate JHK magnitudes from Carter & Meadows (1995), assuming that the stars radiate as black bodies at their effective temperatures. The absolute flux calibration is only measured to a factor of 2 or worse due to the uncertainty in centering the standard stars in the very narrow slit, but the relative flux calibration across the spectrum is well determined. A hot star next to the Crab was observed to determine the profiles of atmospheric absorption bands, which were then removed from the Crab spectra. A final co-added spectrum of each knot was then produced, taking into account the different telescope positions for different on-target exposures. Fig. 1 shows the resulting spectra.

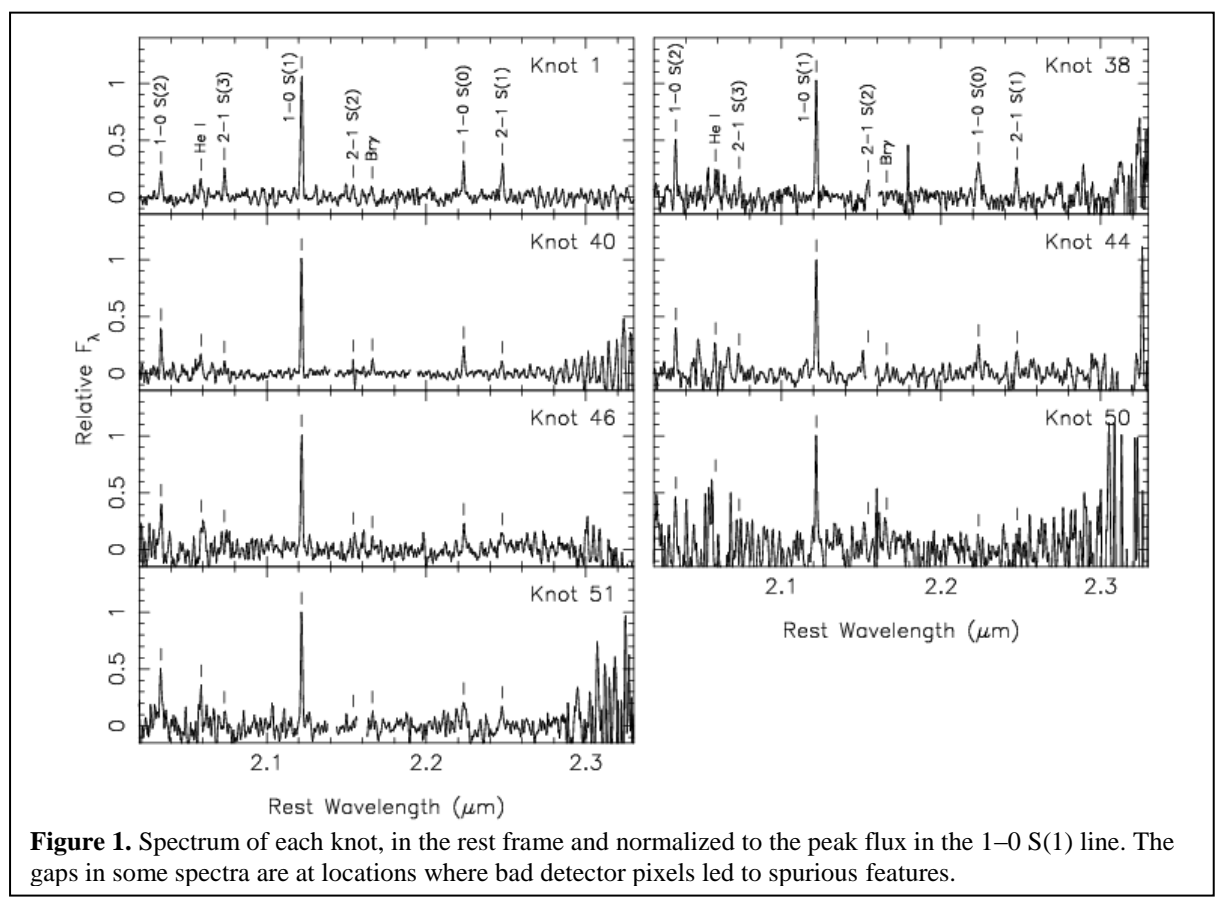

**Figure 1.** Spectrum of each knot, in the rest frame and normalized to the peak flux in the 1–0 S(1) line. The gaps in some spectra are at locations where bad detector pixels led to spurious features.

The final spectrum of each knot shows a well-exposed $H_2$ 2.121µm 1-0 S(1) line and in most cases additional weaker $H_2$ lines. We measured the $H_2$ intensity ratios by fitting the 1-0 S(1) profile to each of the other lines. The results are listed in Table 2. Values labelled '$\leq$' are possible detections, while values labelled '<' are upper limits on undetected lines. The errors in the $H_2$ intensity ratios were estimated from the spectra and scaled by one factor for each knot to make the reduced chi-square equal to 1 for the fit for a single excitation temperature (section 3).

Scaling in this manner reduced the errors by factors in the range 1.2 – 3.3. The table also lists flux measurements or limits for the H I Brγ line. This line is also definitely detected in a region adjacent to but not coincident with Knot 1, following the same velocity structure that is seen for optical emission lines from the FK-10 filament. The He I λ2.059 μm line is also present in several of the extracted spectra, but was not measured.

## 3. The Temperature in the Molecular Region

The $H_2$ temperatures are quite high. This is immediately clear from the fact that we are able to detect lines in the 2–1 vibrational series in 6 of the 7 knots.

In view of the small number of lines measured and the fairly low signal:noise ratio, we assume here that we are measuring a gas in LTE with a single excitation temperature. Relative intensities of $H_2$ lines can then be used to measure that temperature employing level population diagrams such as Fig. 2. We converted the observed $H_2$ intensity ratios into column density ratios through the relation

$$(N_1/g_1) \ / \ (N_2/g_2) = (I_1/I_2) \ (h\nu_2/ \ h\nu_1) \ ( g_2/g_1) \ (A_2/A_1) \qquad (1)$$

where $N_i$ and $I_i$ are the column densities and intensities of the two lines, and $h\nu$, $g$, and $A$ are the photon energy, statistical weight, and transition probability respectively, using the molecular constants summarized in Shaw et al. (2005). Then we assume

$$(N_1/g_1) \ / \ (N_2/g_2) = \exp(-\chi_1/kT_{mol}) \ / \ \exp(-\chi_2/kT_{mol}) \qquad (2)$$

where $T_{mol}$ is the gas kinetic temperature in the molecular region and $\chi$ is the excitation energy of the upper level of the transition. Fig. 2 shows the population ratios we observe as points or upper limits, while the lines give the LTE population ratios at various temperatures. The least-squares best-fitting temperatures are $T_{mol} \sim$ 2200–3200 K, and are listed in Table 2. This measurement of temperature is based primarily on the ratio of the 2-1 S(1) and 1-0 S(1) lines (of ortho-hydrogen) because the errors are much higher for the other lines.

| | Knot 1 | Knot 38 | Knot 40 | Knot 44 | Knot 46 | Knot 50 | Knot 51 |
|---|---|---|---|---|---|---|---|
| | | | | Table 2. Measured parameters | | | |
| $I$(1-0 S(0))/$I$(1-0 S(1)) | 0.23±0.03 | 0.38±0.18 | 0.26±0.04 | 0.28±0.06 | 0.25±0.04 | <0.25 | 0.23±0.04 |
| $I$(1-0 S(1))/$I$(1-0 S(1)) | 1.00±0.03 | 1.00±0.12 | 1.00±0.05 | 1.00±0.07 | 1.00±0.04 | 1.00±0.20 | 1.00±0.04 |
| $I$(1-0 S(2))/$I$(1-0 S(1)) | 0.37±0.05 | 0.51±0.13 | 0.45±0.08 | 0.40±0.11 | 0.41±0.07 | 0.41±0.28 | 0.52±0.09 |
| $I$(2-1 S(1))/$I$(1-0 S(1)) | 0.23±0.02 | 0.34±0.12 | 0.11±0.03 | 0.23±0.07 | 0.19±0.03 | <0.25 | 0.19±0.03 |
| $I$(2-1 S(2))/$I$(1-0 S(1)) | <0.14 | <0.31 | <0.14 | <0.27 | <0.15 | <0.25 | <0.13 |
| $I$(2-1 S(3))/$I$(1-0 S(1)) | 0.34±0.05 | ≤0.17 | <0.14 | <0.19 | <0.17 | <0.41 | <0.28 |
| $I$(Brγ)/$I$(1-0 S(1)) | <0.09 | <0.09 | 0.13±0.04 | <0.13 | ≤0.16 | <0.24 | <0.13 |
| $T_{mol}$ (K) | 3198±66 | 2944±146 | 2221±540 | 3065±514 | 2837±207 | ≤2753 | 2837±232 |
| $v_{helio}$(H$_2$) (km s$^{-1}$) | 145 | 390 | 204 | 737 | -540 | 123 | 108 |
| $v_{helio}$([S II]) (km s$^{-1}$) | 90 | 371 | 188 | 737 | -525 | 128 | 146 |
| $n_e$([S II]) (cm$^{-3}$) | 1500 | 2100 | 1400 | 1400 | 2400 | 2500 | 2100 |
| $n_{mol}$ (cm$^{-3}$) | 14000 | 21000 | 19000 | 14000 | 25000 | ≤28000 | 22000 |
| $l_{los}$ (10$^{13}$ cm) | 25. | 7.4 | 14. | 11. | 5.4 | 3.9 | 8.0 |
| $l_{los}$ / $width$ | 0.0018 | 0.0013 | 0.0010 | 0.0010 | 0.0007 | 0.0006 | 0.0012 |
| Ortho:para ratio | 3.0±0.5 | 2.3±0.5 | 2.46±0.16 | 2.6±0.6 | 2.7±0.4 | --- | 2.44±0.20 |

There may be some evidence that the temperatures vary with the vibrational level. For Knot 1, the 1–0 lines (of para-hydrogen) appear to correspond to temperatures in the 1500–2000K range, whereas the 2–1 lines (of ortho-hydrogen) suggest $T_{mol} >$ 4000K (see Fig. 2). Knot 1 is the only one with two well-measured 2–1 lines. For Knots 40 and 51, the temperature of the 1–0 lines agree with the temperature derived from the 2–1 S(1) and 1–0 S(1) lines. For Knot 46, the 1–0 lines appear to correspond to temperatures in the 1500–2000K range. Knots 38, 44, and 50 have insufficient measurements to derive a temperature for the 1–0 or 2–1 level.

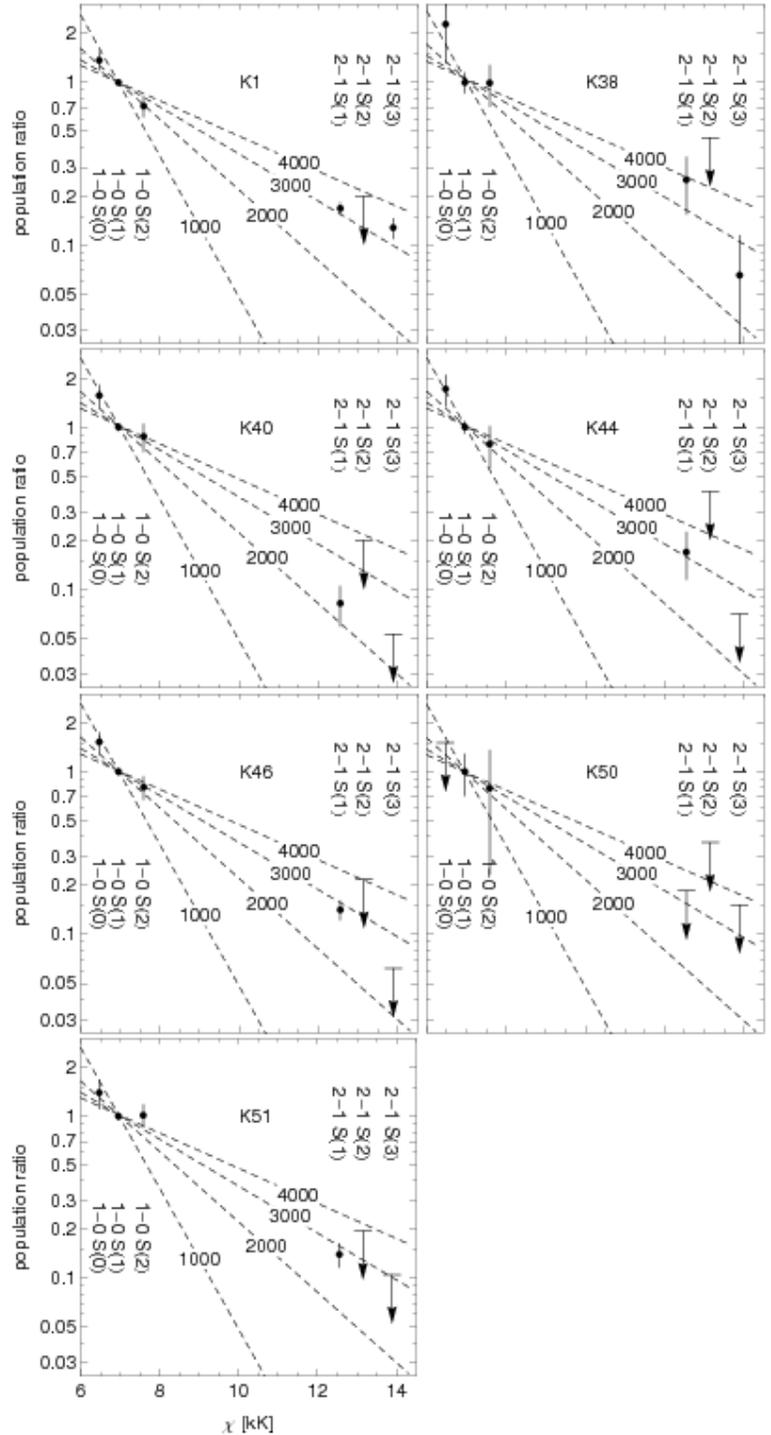

**Figure 2.** Measured populations of the upper levels producing various H$_2$ lines divided by the number of nuclear spin and rotational states and normalized to the 1-0 S(1) line. The horizontal axis is $\chi$, the excitation energy for the upper level. The dashed lines are the predicted relative populations corresponding to several different temperatures. These lines have been forced to pass through the point for the fiducial 1–0 S(1) emission line, although we included error bars for this line in our least-squares fits.

## 4. The Ortho-to-Para Ratio

In the population plots (Fig. 2), the lines involving transitions of ortho-hydrogen are offset from those of para-hydrogen. The points in the figure were made by dividing by the multiplicity, which includes the nuclear multiplicity, 3 for ortho-hydrogen and 1 for para-hydrogen. For example, consider Knot 51: the lines of para-hydrogen are high in the figure, which, taken at face value, means the ortho-to-para ratio is less than the ratio of multiplicities.

To find the ortho-to-para ratio, we compared the 1-0 S(0) and 1-0 S(2) lines of para-hydrogen with the 1-0 S(1) line of ortho-hydrogen, and we assumed that the gas is in local thermal equilibrium and that the all of the gas is at the same temperature. We scaled the populations according to the Boltzmann factors for the temperature found by fitting all of the lines ($T_{mol}$ in Table 2). ($T_{mol}$ was derived primarily from the ratio of the lines with high signal-to-noise ratios, the 1-0 S(1) and 2-1 S(1) lines, which are from two different vibrational levels.) The measured abundance ratio of ortho:para is 3 times the weighted mean of the two scaled ratios.

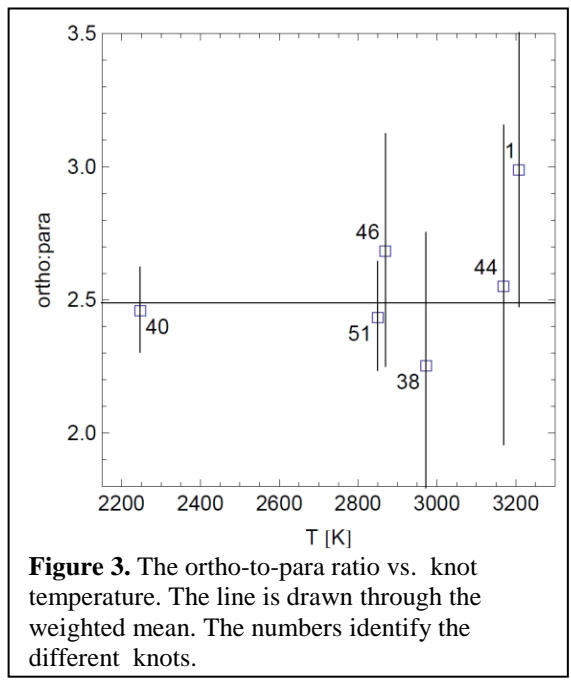

**Figure 3.** The ortho-to-para ratio vs. knot temperature. The line is drawn through the weighted mean. The numbers identify the different knots.

The ortho-to-para ratios are listed in the last row of Table 2 and shown in Fig. 3. The weighted mean ortho-to-para ratio is 2.5. This result should be read with caution. With so few lines, we are unable to test our assumptions. Additional observations are needed. Observations of the 1–0 S(3) line would be extremely useful, since it would be a second ortho line with which one can get a second measurement of the ortho-to-para ratio. With spectra having better signal-to-noise ratio, it would be possible to determine the rotational temperature of the ν=1 vibrational state. Finally, we have assumed local thermal equilibrium, but whether that is sensible must be tested with a plasma simulation model.

## 5. Density Estimates

The gas density of H in any form in the molecular region, $n_{mol}$, can be estimated by assuming gas pressure balance with a surrounding layer of ionized gas. We showed in Paper II that the $H_2$ knots are associated with [S II]-emitting gas along the same line of sight. Here we find that the radial velocities of the $H_2$ 2.121µm lines agree well with the [S II] velocities reported in Paper II (compare rows 9 and 10 of Table 2 in this paper). This strengthens the case that the bright [S II] features really are physically associated with the $H_2$ knots in the expanding three-dimensional structure of the Crab.

We used the optical spectra described in Paper II to measure the [S II] λ6731/λ6716 intensity ratio in each knot and from it found the electron densities $n_e$ listed in row 11 of Table 2. We then computed $n_{mol} = 2\,(T_{rec}/T_{mol})\,n_e$, where $T_{rec} \sim 15{,}000$K (appropriate for a Crab filament) is the kinetic temperature in the recombination (ionized) zone. The factor of 2 accounts for the fact that in the recombination region the total particle density is about twice the measured electron density (assuming that He is singly ionized and that the He/H abundance ratio is 0.5), while in the molecular region the total density of H baryons is about two times the $H_2$ particle density, but about equal to the total particle density including the free helium atoms. The resulting values of $n_{mol}$ are listed in row 12 of Table 2.

## 6. Discussion

### 6.1 The structure of the Crab filaments

Fig. 4 shows the $H_2$ 1–0 S(1) $\lambda 2.121\mu m$ emissivity per particle (i.e. per H baryon), $4\pi j / n_{mol}$, as a function of the temperature $T_{mol}$ and the H baryon density $n_{mol}$. A similar figure, but showing $4\pi j / n_{mol}^2$, was discussed in some detail in Paper I. The CLOUDY model (Ferland et al. 1998) and the physics assumed for the figures are from Ferland et al. (2009) for the case of "ionizing particles", but the results are not sensitive to the source of the energy input. The 0–1 S(1) emissivity per particle is low at low temperatures because the high levels producing the observed $H_2$ lines are not excited, and the cutoff at temperatures above $\log T_{mol} \sim 3.5$ is due to collisional dissociation of the molecule. The positions of the filled circles on Fig. 4 show the observed values. They fall very near the peak emissivity per particle. This is almost certainly the selection effect discussed in Paper I. There is likely to be additional $H_2$ at temperatures too low for it to be detectable, and $H^0$ that has been produced by the dissociation of $H_2$. It also means that the $H_2$ that we do detect has somehow been heated nearly to its dissociation temperature.

The observed points on Fig. 4 cluster at $n_{mol} \sim 2\times10^4$ cm$^{-3}$, at the low-density edge of the region where the predicted emissivity per particle no longer depends on density because the populations of the upper levels involved in the 2.12 μm transition have come into LTE. There could be additional molecular material at lower densities that is actively emitting but is too faint to be detectable.

The $n_{mol}$ values found here are based on the constant gas pressure assumption together with the observed density and temperature in the $H^+$ region. The slanting solid line on Fig. 4 shows the locus of such points. However, these are really *lower* limits to the actual gas density because the molecular gas might be confined by magnetic pressure or because gas flowing off the surface of a dense molecular cloud would produce a strong density gradient across the $H^0$ and $H^+$ zones because of mass conservation.

At the density corresponding to gas-pressure equilibrium, the column density through a typical 3 arcsec $\sim 10^{17}$ cm projected width on the sky of an $H_2$ knot is $N_{mol} \sim 10^{21}$ cm$^{-2}$. It can easily be shown that $H_2$ cannot survive destruction from the Crab's synchrotron-continuum within such a small column density. Fast processes such as H⁻ back neutralization (see Doughty, Fraser & McEachran 1966) set a lifetime of just a few days for $H_2$ molecules directly exposed to the continuum radiation. We used our plasma simulation code CLOUDY (Ferland et al. 1998) with ISM grains and other input parameters taken from Pequignot & Dennefeld (1983) to make a constant pressure model of gas exposed to the Crab's synchrotron radiation and with the observed electron density in the $S^+$ zone, and find that the $H_2$ region does not form until the total H column density is $N_H \sim 10^{22}$ cm$^{-2}$. This all shows that well-shielded layers of $H_2$ with much higher gas density *must* exist. This is a critical distinction between SNR filaments and the cool-core galaxy cluster filaments, which are exposed to very feeble radiation (Ferland et al. 2009).

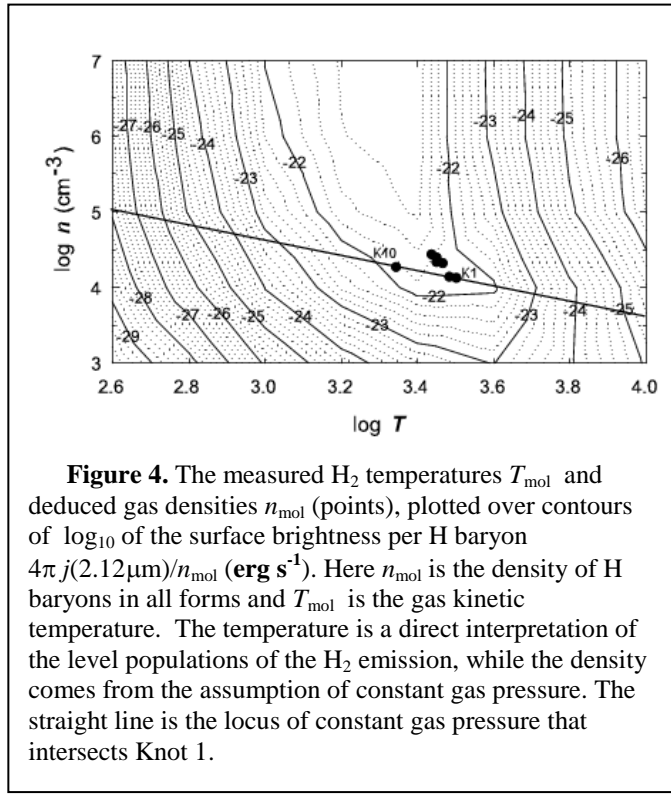

**Figure 4.** The measured $H_2$ temperatures $T_{mol}$ and deduced gas densities $n_{mol}$ (points), plotted over contours of $\log_{10}$ of the surface brightness per H baryon $4\pi j(2.12\mu m)/n_{mol}$ (**erg s$^{-1}$**). Here $n_{mol}$ is the density of H baryons in all forms and $T_{mol}$ is the gas kinetic temperature. The temperature is a direct interpretation of the level populations of the $H_2$ emission, while the density comes from the assumption of constant gas pressure. The straight line is the locus of constant gas pressure that intersects Knot 1.

The line-of-sight thickness $l_{los}$ of the molecular region that contributes to the observed $H_2$ emission can be estimated from the ratio of the $H_2$ 0–1 S(0) intensity to the emissivity per unit volume. For each knot, we determine the intensity $4\pi J(H_2)$ in the $\lambda 2.121$ μm line from the average $H_2$ surface brightness $S_{avg}(H_2)$ that was measured using narrow-band direct imaging (Paper II). The emissivity per unit volume $4\pi j(H_2)$ is given by Fig. 3 (after multiplying the value in the figure by $n_{mol}$). The resulting thicknesses are $l_{los} \sim 10^{14}$ cm, and are listed for each knot in Table 2. These thicknesses are only about a fraction $10^{-3}$ of the projected widths of the knots on the sky, and are actually upper limits because the emissivity $4\pi j(H_2)$ will be higher if the density is higher than our estimate of $n_{mol}$.

If the $H_2$ knots are monolithic clumps that are roughly spherical, the $10^{-3}$ thickness-to-width ratios mean that the knots would have very large non-emitting cores, so the total mass of $H_2$ in each knot would be about 1000 times greater than the mass we can detect. In Paper I we estimated the mass of Knot 1 to be $M_{mol} \sim 5 \times 10^{-5}$ $M_\odot$ for $n_{mol} = 10^4$ cm$^{-3}$. A factor of 1000 correction would suggest about 0.05 $M_\odot$ of molecules in Knot 1, and a minimum total of about 1 $M_\odot$ of $H_2$ in the approximately 50 knots that we have observed. Since the temperature in the non-emitting interior is cooler (Temim et al. 2006 found dust at about 80K), their density is likely to be substantially greater than $10^4$ cm$^{-3}$. Therefore the mass of $H_2$ may be significantly greater than 1 $M_\odot$. In this scenario, dust absorption in the cool core would cause knots like Knot 1 to appear black if seen in silhouette.

An opposite extreme would be for the very small thickness-to-width ratio to come about because each 3 arcsec $H_2$ knot seen on the sky is actually composed of a fine mist of very dense molecular concentrations. The volume filling factor would then be $f \leq 10^{-9}$. For $n_{mol} > 10^4$ cm$^{-3}$, the emissivity per particle is constant, so the total mass in all concentrations needed to produce the observed $\lambda 2.121$ μm emission from Knot 1 would be $M_{mol} \sim 5 \times 10^{-5}$ $M_\odot$ and for all 40 knots would be $2 \times 10^{-3}$ $M_\odot$ – the extra factor of 1000 is no longer needed because all of the $H_2$ is now assumed to be actively emitting. In this case, a structure like Knot 1 would be very porous to background radiation, so would not necessarily produce much dust absorption of the synchrotron emission coming from behind the knot. The CO spectrum would be very different for these two cases.

A complete plasma simulation model of a filament may reveal other possibilities besides the highly molecular situation discussed above, but we do not yet have all of the necessary data in hand to construct such a model. In particular, CO measurements for one or a few individual filaments would be of tremendous value.

## 6.2 The origin of the H$_2$ emission

Understanding the origin of the strong H$_2$ emission is fundamental if we are to use the H$_2$ lines to probe conditions in the gas. We are still in the process of gathering the needed observational database to fully describe what is happening in molecular regions in the Crab filaments. We have assumed that the H$_2$ lines are collisionally excited. Our excitation diagrams are then temperature indicators if LTE holds. Collisional excitation occurs if the gas is heated to warm temperatures, perhaps by shocks or energetic particles. Crab filaments are morphologically similar to those seen in cool-core galaxy clusters where similar high ro-vibration H$_2$ temperatures are found (Hatch et al. 2005; Johnstone et al. 2007). In the case of the cluster filaments, we have shown that the emission is produced by ionizing particles (Ferland et al. 2009).

UV fluorescence, rather than collisional excitation, produces the H$_2$ emission from PDRs near H II regions (Tielens 2005; Osterbrock & Ferland 2006; Draine 2011; Ferland 2011). Photons around 0.1 μm are absorbed into H$_2$ electronic excited configurations. Roughly 90% then relax to populate excited levels within the ground electronic configuration (Tielens & Hollenbach 1985). This produces faint H$_2$ lines with a highly non-thermal level population distribution (Black and van Dishoeck 1987; Sternberg & Dalgarno 1989). Sternberg & Neufeld (1999) show that UV fluorescence also produces a low OP ratio due to stronger self-shielding in the ortho levels. Could UV fluorescence be responsible for the strong H$_2$ emission, the high excitation temperature, and the possibly low OP ratio that we find?

Empirically, UV fluorescence produces H$_2$ lines that are a tiny fraction of the nearby H I recombination lines in H II regions. Table 2 in Paper I shows that the H$_2$ / Brγ intensity ratio is 0.07 for the Orion Bar. Table 2 of this current paper shows that the measured H$_2$ / Brγ intensity ratios are 6 for one knot and lower limits of 5-11 for the others. It is likely that the ratio is higher yet in some knots, since Brγ and H$_2$ are not correlated spatially in the vicinity of most of the H$_2$ knots (Paper II).

Simple photon-counting arguments show that UV fluorescence contributes very little to the observed H$_2$ emission in the Crab. We estimate the largest possible H$_2$ / H I ratio. We can obtain an upper limit to the H$_2$ emission if we assume that all continuum photons around 0.1 μm are absorbed by H$_2$ and that 90% then decay into X, the ground electronic configuration. H I recombination lines are produced by absorption of the radiation field shortward of 0.0912 μm. The H$_2$ / H I ratio is then

$$\frac{I(\text{H}_2)}{I(\text{H I})} \leq k \frac{\phi(0.0912 \mu m - 0.11 \mu m)}{\phi(\text{H I})} \qquad (3)$$

where $\phi$(H I) is the flux of hydrogen ionizing photons in the SED and $\phi$(0.0912 μm – 0.11 μm) is the flux in the 0.0912 – 0.11 μm band. The proportionality constant $k$ depends on which lines are considered.

This equation assumes that all photons in each of these bands are used to produce either H I or H$_2$ lines. The H I emission will be overestimated if there is no H$^+$ ionization front, but there will be no H$_2$ emission at all in this case. Observationally we know that the H$_2$ core is surrounded by an ionized region with lines of neutral species formed between the ionized and H$_2$ zones (cf. Paper I; Paper II; Sankrit et al. 1998), so an H$^+$ ionization front is clearly present. The H$_2$ emission will be overestimated if there is not sufficient column density for $H_2$ to form or if dust absorbs the radiation field.

The flux ratio in equation 3 is obtained by comparing the SED of the Crab Nebula (Davidson & Fesen 1985; Hester 2008) with the Atlas O star atmosphere used by Baldwin et al. (1991) in their model of the Orion H II region. Figure 5 shows both SEDs. The two SEDs are normalized to have the same intensity at 0.1 μm, the wavelength where fluorescence is active, and so will produce the same $H_2$ emission. This $H_2$ emission is to be compared with the intensities of H I recombination lines, which are produced by photons shortward of 0.0912 μm. Numerical integration verifies what the eye sees – there are 1 dex more H-ionizing photons in the Crab SED than in the O star. The ratio given in equation 3 is 1 dex smaller than in the Orion Nebula. It follows that UV fluorescence photoexcitation and H I recombination will produce a ~1 dex smaller $H_2$ / H I intensity ratio in the Crab than in Orion, which is itself about 2 dex smaller than that seen in the Crab. This means that less than about $10^{-3}$ of the observed $H_2$ emission in the Crab filaments originates from UV fluorescence. The ineffectiveness of UV fluorescence has previously been pointed out by G90 and in Paper I. There simply are not enough photons around 0.1 μm to produce the observed $H_2$ lines.

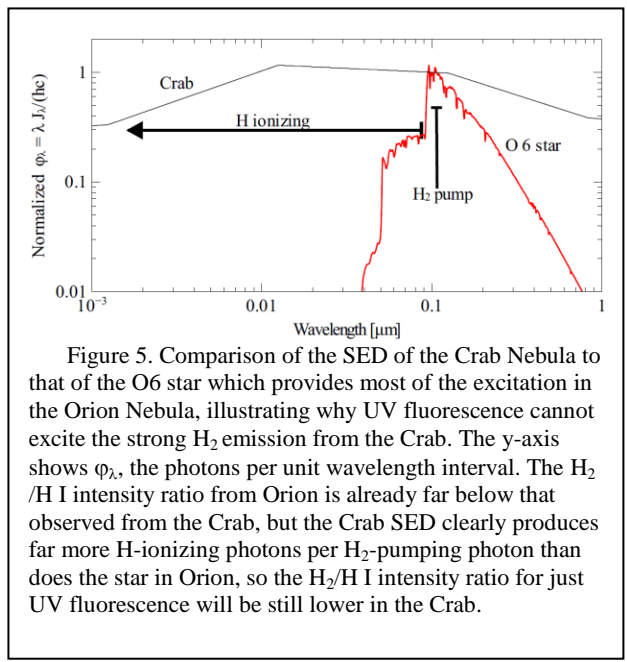

Figure 5. Comparison of the SED of the Crab Nebula to that of the O6 star which provides most of the excitation in the Orion Nebula, illustrating why UV fluorescence cannot excite the strong $H_2$ emission from the Crab. The y-axis shows $\varphi_\lambda$, the photons per unit wavelength interval. The $H_2$/H I intensity ratio from Orion is already far below that observed from the Crab, but the Crab SED clearly produces far more H-ionizing photons per $H_2$-pumping photon than does the star in Orion, so the $H_2$/H I intensity ratio for just UV fluorescence will be still lower in the Crab.

It is true that UV fluorescence can produce a distribution of level populations that indicates an apparent (and false) high temperature and an OP ratio less than three, and such a model could in principal be fitted to our $H_2$ data. However, it is not correct to invert this logic and say that high rotation-vibration temperatures, or low OP ratios, show that the $H_2$ lines are produced by UV fluorescence. Energy conservation arguments like those given above take precedence, especially since other excitation processes are possible.

We quantify the photon problem in the Crab in Table 3. This considers four well observed examples: the Orion Bar, the average of 8 PNe for which Davis et al. (2003) list both $H_2$/Brγ and OP ratios, the Crab Nebula ($H_2$ Knot 51, an isolated knot, which is free of confusion from multiple filaments seen in projection), and the Horseshoe region of the Perseus cool-core galaxy cluster (Hatch et al. 2005; Johnstone et al. 2007). Column 2 lists the ratio of the number of photons between 1100A and 912A, which Bertoldi & Draine 1996 use as a fundamental PDR radiation parameter, to the number of photons shortward of 912A. There is no entry for the Horseshoe since starlight is not seen there. For the PNe we use a $10^5$ K Rauch (1997; 2002) atmosphere. The sources for the Orion and Crab SEDs are the same as for Figure 5. The last column of the table gives the observed ratio of the $H_2$ 2.121 micron line to H I Brγ. The value for knot 51 is from Paper II, which derived this ratio from narrow-band images. The Orion Bar serves as a baseline for UV fluorescence – photon pumping is thought to account for $H_2$ emission across the Bar. For the PNe and the Crab knot, the $H_2$/HI ratio expected from only UV fluorescence will scale from the Orion values by the ratio of φ(H$_2$)/φ(H), and must be extremely small (0.003 and 0.008, respectively).

| Table 3. Comparison of SED to H$_2$/H I Line Strength Ratio | | | |
|---|---|---|---|
| Object | φ(H$_2$)/φ(H I) | Scaled I(H$_2$ 2.12μm)/I(H I Brγ) for UV fluorescence | Observed I(H$_2$ 2.12μm)/I(H I Brγ) |
| Orion Bar | 1.83 | 0.07 | 0.07 |
| PNe | 0.08 | 0.003 | 4 |
| Crab H$_2$ knot | 0.22 | 0.008 | 10 |
| Cool-core cluster filament | --- | --- | 8 |

Understanding the molecular gas in detail hopefully would tell us where it originated and its history. There are two competing models for the filaments, and a better determination of the OP ratio might offer an important clue as to which (if either) is correct. Hester et al. (1996) argue that gas, originally in an unseen highly-ionized outer shell, is overtaken by the expanding synchrotron bubble. The hot gas is compressed and, through various MHD instabilities, flows down the filament and is deposited in the 'head', which includes the observed H$_2$ knot. In this scenario the H$_2$ forms from warm ionized gas and we would expect an OP ratio close to 3. Another possibility is that the 'head' of the filament is a dense clump that existed within the atmosphere of the SN progenitor (Carlqvist 2004). The filament is a comet-tail like structure that is swept away from the molecular core by the expanding ejecta. In this case the OP ratio of the core might reflect conditions in the condensations that preceeded the explosion. It seems more likely that molecular gas formed with a low OP ratio in colder conditions would be present in the latter rather than the former model.

### 6.3 Implications for cool-core cluster filaments

The H$_2$ excitation temperatures found here are surprisingly similar to those found in cool-core cluster filaments (Jaffe, Brehmer & van der Werf 2001), and are vastly warmer than is found in the H$_2$ regions associated with the H II regions and PDRs found around young star clusters. In the case of the cluster filaments the H$_2$ excitation temperature increases with increasing excitation potential (Johnstone et al. 2007), a sign of a range of kinetic temperatures extending from very low to high values (Ferland et al. 2008). The interpretation is that ionizing particles, the hot gas surrounding the filaments, penetrates cold molecular regions in the filaments, which then expand along magnetic field lines to maintain constant gas pressure. This accounts for all molecular, atomic, and ionic emission lines seen in those filaments.

The filaments in the Crab are similar in many ways. Clearly there are ionizing particles present – these produce the synchrotron emission. A distinction is that the Crab filaments are in a photon-rich environment – the Crab is bright across much of the electromagnetic spectrum. The synchrotron continuum is produced by the ionizing particles, so the two may be strongly correlated. It may be quite difficult to disentangle which is the prime mover in energizing the molecular cores. Clearly more observations are needed.

## 7. Conclusions

We have measured the $H_2$ excitation temperature in six of the seven molecular knots that we observed, and find $T_{mol}$ = 2000–3000 K. This is near the peak $H_2$ line emissivity shown in Paper I, suggesting that we are selectively seeing only the $H_2$ with the highest emissivity per particle and that extensive non-emissive regions could also exist. In Paper II we showed that there are also much more extensive regions of low-surface-brightness $H_2$ emission, which could be the tip of an iceberg of cold molecular gas.

We have measured the density in ionized gas that is associated with the molecular knots. We then estimated the density in the molecular gas assuming that the regions are in gas pressure equilibrium. Since there may be some other means of confining the molecular knots, this establishes a lower density limit in the molecular region of $n_{mol} \geq 14,000$ cm$^{-3}$.

The corresponding depth of the $H_2$-emitting layer is only $10^{-3}$ of its width, or less. This suggests that extensive amounts of $H_2$ that is too cold to emit are also present within the observed knots. In this picture, the total mass in the approximately 40 observed, bright $H_2$ knots would be about 1 $M_\odot$ or higher. This cold $H_2$ could be detected by its CO emission. An alternative picture is that the molecular gas forms a fine mist within the confines of the $H_2$ emission knot seen on the sky. In this case there could be very little non-emitting $H_2$ within the bright molecular knots. Future CO observations could tell the difference between these two cases.

The actual excitation mechanism remains unclear, but UV fluorescence is not a significant contributor. Penetrating cosmic rays seem likely, and would strengthen the analogy with the filaments in cool core clusters of galaxies (Ferland et al. 2009).

These current results emphasize the many similarities between the Crab Nebula's filaments and the far-larger filamentary structures that have formed in the IGM of cool-core galaxy clusters. Both environments are permeated by strong fluxes of energetic particles which may play a major role in heating the molecular gas. The Crab is the nearby laboratory for understanding the distant cluster filaments.

*Acknowledgments.* We thank Bob O'Dell for helpful comments on the manuscript. EDL, JAB and ZKC are grateful for support from NASA through ADP grant NNX10AC93G. GJF acknowledges support from NSF (0908877), NASA (07-ATFP07-0124 10-ATP10-0053, and 10-ADAP10-0073) and STScI (HST-AR-12125.01 and HST-GO-12309).